\title {\Large\bfseries
			Une m\'ethode Monte-Carlo pour les m\'eandres \\
                        sur ordinateur parall\`ele 
}
\author { 		O. Golinelli
\\ \ad			Cea Saclay, Service de physique th\'eorique,
\\ \ad			91191 Gif-sur-Yvette, France
\\ \ad			email: golinelli@cea.fr
}
\date{\normalsize	sept. 1999
\\			Preprint T99/109   ; cond-mat/9910004
}

\documentclass[
  a4paper, 
  12pt, onecolumn,
  draft
]{article}

\usepackage{epsf}		            % pour inclure les figures *.eps
\usepackage[french]{babel}	            % style francais

\newcommand  {\ad}{\normalsize\em}	    % style de l'adresse

\newcommand  {\largeurFig} {8.5cm}	    % largeur maxi des figures

\begin{document}
\maketitle

%%%%%%%%%%%%%%%%%%
\begin{abstract}
%%%%%%%%%%%%%%%%%%

Pour mod\'eliser le repliement compact d'un polym\`ere, on \'etudie la
statistique des m\'eandres, d\'efinis comme configurations d'un circuit
automobile traversant $n$ fois une rivi\`ere.  Gr\^ace \`a une m\'ethode
Monte-Carlo adapt\'ee \`a un ordinateur massivement parall\`ele, les
m\'eandres sont simul\'es jusqu'\`a la taille $n=400$.  Le comportement
asymptotique du d\'enombrement et de l'enroulement moyen des m\'eandres
sont pr\'esent\'es.

\medskip \noindent Mots-clef: repliement, pliage, m\'eandres, Monte-Carlo,
arbre, parall\'e\-lisme

\end{abstract}

%%%%%%%%%%%%%%%%%%%%%%%%
\section{Introduction}
%%%%%%%%%%%%%%%%%%%%%%%%

Un des centres d'\'etude actuel de la m\'ecanique statistique est le
repliement d'objets g\'eom\'etriques qui fluctuent, comme les membranes
(\`a 2 dimensions) ou les polym\`eres (\`a une dimension).  Les exemples en
physique et en biologie sont nombreux.  Les concepts d\'evelopp\'es pour
les ph\'enom\`enes critiques (observ\'es lors de certaines transitions de
phase) s'appliquent aussi \`a ce domaine.  En particulier, avec
``l'universalit\'e'', certaines quantit\'es d\'ecrivant le comportement
critique \`a grande \'echelle sont ind\'ependantes des d\'etails \`a petite
\'echelle (forme des monom\`eres, interactions, \ldots).  Le physicien
th\'eoricien cherche alors \`a repr\'esenter un objet complexe (par
exemple, un polym\`ere biologique) par le mod\`ele, le plus simple possible
pour pouvoir faire des calculs, mais dont on pense qu'il pr\'esente les
m\^emes propri\'et\'es physiques \`a grande \'echelle.

Dans cet article, le repliement compact d'un polym\`ere est mod\'elis\'e
par une bande de timbres-poste compl\`etement repli\'ee~\cite{saintelague}.
Ceci est \'equivalent au probl\`eme des {\em m\'eandres}, r\'esum\'e par
cette simple question~: de combien de mani\`eres $M_n$ un circuit
automobile peut-il traverser une rivi\`ere en $n$ ponts, en contournant
\'eventuellement la source~?

Une variante~\cite{lando_zvonkin}, o\`u on interdit le contournement de la
source, est \'equi\-valente \`a l'\'enum\'eration des labyrinthes d'un
certain type~\cite{phillips}, ainsi qu'au 16i\`eme probl\`eme de Hilbert,
c'est \`a dire l'\'enum\'eration des ovales de courbes planaires
alg\'ebriques~\cite{arnold}.

Ce probl\`eme a par ailleurs des connections avec la th\'eorie des
n{\oe}uds, les ``mod\`eles de matrice'' et la ``gravit\'e quantique'', et
m\^eme la QCD (th\'eorie des interactions nucl\'eaires fortes en physique
des particules).

Les m\'eandres sont donc un mod\`ele de repliement compact d'un objet \`a
une dimension, avec des maillons identiques, dont on ne retient que l'ordre
dans le pliage.  Malgr\'e cette simplification extr\^eme, ce probl\`eme
r\'esiste depuis au moins un si\`ecle~: par exemple, on ne conna\^{\i}t
toujours pas de formule pour le nombre de pliages possibles $M_n$, ni
m\^eme son comportement asymptotique pour $n$ grand.

L'approche pr\'esent\'ee ici est purement num\'erique.  Elle a \'et\'e
rendue possible en exploitant la puissance des ordinateurs parall\`eles.
Cependant, la m\'ethode de ``brute force'' ne permet pas d'aller bien loin
et un algorithme Monte-Carlo efficace a \'et\'e \'elabor\'e.  La nature
m\^eme des m\'eandres rend difficile la vectorisation des boucles les plus
internes du programme.  Par contre, celui a \'et\'e parall\'elis\'e avec
des ``gros grains'', c'est \`a dire en divisant le travail en sous-t\^aches
au plus haut niveau possible.

 La Section 2 est consacr\'ee aux d\'efinitions.  La Section~3 pr\'esente un
algorithme de d\'enombrement exact, lui-aussi parall\'elis\'e.  La Section~4
d\'ecrit la m\'ethode Monte-Carlo.  Finalement, quelques r\'esultats sont
expos\'es Section~5.

%%%%%%%%%%%%%%%%%%%%%%%%%
\section{D\'efinitions}
%%%%%%%%%%%%%%%%%%%%%%%%%

Un {\em m\'eandre} de taille $n$ est d\'efini comme une configuration
planaire d'une boucle qui ne se croise pas elle-m\^eme (la {\em route}),
croisant une demi-droite (la {\em rivi\`ere} avec sa {\em source}) en $n$
points (les {\em ponts}).  Deux m\'eandres de m\^eme taille sont
consid\'er\'es comme \'equivalents si l'un est obtenu \`a partir de l'autre
en d\'eformant contin\^ument la route tout en gardant fix\'es les ponts.
C'est une \'equivalence topologique.

On appelle {\em arche} chaque section de route entre deux ponts.  Un
m\'eandre de taille $n$ a donc $n$ ponts et $n$ arches.

\begin{figure}
  \centering\leavevmode
  \epsfbox{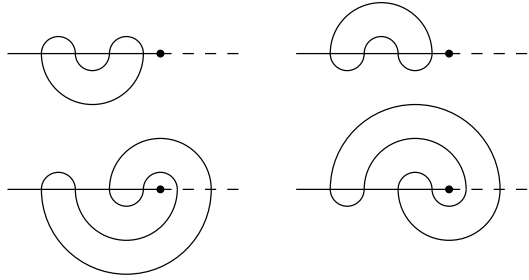}
  \caption{\em
    Les $M_4=4$ m\'eandres de taille $4$.  La {\em route} est la boucle
    auto-\'evitante.  La {\em rivi\`ere} semi-infinie est la
    demi-droite en trait plein, commen\c{c}ant \`a la {\em source}
    (point noir).  La {\em taille} est le nombre de ponts.  Le nombre
    d'enroulements $w$ est le nombre d'arches croisant la demi-droite
    en pointill\'e \`a droite de la source.  Les deux m\'eandres du haut
    n'ont pas d'enroulement ($w=0$), mais les deux du bas ont $w=2$.
  }
  \label{m4}
\end{figure}

Le nombre de m\'eandres diff\'erents de taille $n$ est not\'e $M_n$.  Par
exemple, $M_1=1$, $M_2=1$, $M_3=2$ et $M_4=4$.  Sur la Fig.~\ref{m4}, les 4
m\'eandres de taille 4 sont dessin\'es.

\begin{figure}
  \centering\leavevmode
  \epsfxsize=\largeurFig
  \epsfbox{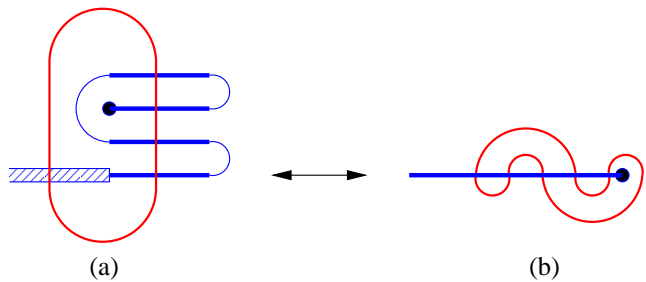}
  \caption{\em
    On transforme un pliage de $n$ timbres (ici $n=4$) en un m\'eandre
    de taille $n+1$ de la mani\`ere suivante : (a) dessiner une boucle ovale
    traversant la pile de $n$ timbres et se refermant sur la gauche en
    traversant le support.  Il y a donc $n+1$ intersections. (b) Mettre
    \`a plat la bande de timbres en tirant vers la droite son
    extr\'emit\'e.  On obtient alors un m\'eandre avec $n+1$ ponts, la
    boucle devenant la route, et la bande de timbres la rivi\`ere.  La
    transformation inverse donne, \`a partir de tout m\'eandre,
    une bande de timbre totalement repli\'ee, en d\'eformant la
    route pour en faire une boucle ovale.
  }
  \label{timbres}
\end{figure}

Le probl\`eme des m\'eandres est absolument \'equivalent \`a celui du
repliement d'une bande de timbres-poste.  Consid\'erons une bande de
$n$ timbres attach\'ee \`a un support.  Il y a alors $M_{n+1}$
mani\`eres de plier cette bande compl\`etement, c'est \`a dire sous
forme de pile avec un seul timbre en largeur.  En effet, comme
expliqu\'e Fig.~\ref{timbres}, chaque pliage de $n$ timbres correspond
\`a un m\'eandre de taille $n+1$ et r\'eciproquement.  On pr\'ef\`ere
cependant utiliser dans cet article la repr\'esentation sous forme de
m\'eandre car elle est plus naturelle pour d\'ecrire la r\'ecurrence avec
laquelle nous les simulerons.

Les m\'eandres ont des similitudes avec d'autres probl\`emes de
physique statistique, par exemple les marches al\'eatoires
auto-\'evitantes ferm\'ees~: un m\'eandre est obtenu \`a partir d'une
telle marche en lui superposant une demi-droite et en ne gardant que
l'aspect topologique, c'est \`a dire la nature des intersections.
Aussi, par analogie, on attend que
  \begin{equation}
    M_n \stackrel{n \rightarrow \infty}{\sim} c \; {R^n \over n^\gamma},
    \label{rnng}
  \end{equation}
o\`u les estimations~\cite{lunnon,dfgg1} sont $R \sim 3.5$ et $\gamma \sim
2$.

Le nombre $R$, \'equivalent de la connectivit\'e, peut \^etre
interpr\'et\'e comme le nombre moyen de possibilit\'es pour ajouter un pont
pr\`es de la source en ne d\'eformant qu'une seule arche, pour un grand
m\'eandre.  Ainsi, $\ln(R)$ appara\^{\i}t comme l'entropie par pont~; c'est
une quantit\'e intensive, c'est \`a dire d\'efinie par unit\'e de volume.
Par contre, l'exposant $\gamma$ sera sensible aux conditions au bord.  Par
exemple, si l'on change la d\'efinition du probl\`eme en prenant une
rivi\`ere infinie aux deux bouts, ou bien une rivi\`ere avec un affluent,
alors la connectivit\'e $R$ ne sera pas chang\'ee (elle caract\'erise
l'effet du volume), mais par contre $\gamma$ sera \`a chaque fois
diff\'erent.  Num\'eriquement, $\gamma$ est plus difficile \`a mesurer car
il d\'ecrit la correction sous-dominante dans l'\'equation~(\ref{rnng}).

Pour un m\'eandre donn\'e, le nombre d'{\sl enroulements} de la route
autour de la source est d\'efini comme le nombre minimal d'intersections
entre la route et une demi-droite (semi-infinie) partant de la source en
prolongeant la rivi\`ere du cot\'e oppos\'e.  On peut le voir comme la
distance topologique bout-\`a-bout entre la source (extr\'emit\'e droite de
la rivi\`ere) et l'infini (extr\'emit\'e gauche), la distance entre deux
points \'etant d\'efinie comme le nombre minimal d'arches \`a traverser
pour aller d'un point \`a l'autre, sans franchir la rivi\`ere.  Pour un
exemple, voir Fig.~\ref{m4}.

On d\'efinit $w_n$ comme la moyenne du nombre d'enroulements sur tous les
$M_n$ m\'eandres de taille $n$.  Par analogie avec les marches
al\'eatoires, on s'attend \`a
  \begin{equation}
    w_n \stackrel{n \rightarrow \infty}{\sim}  n^{\nu},
  \end{equation}
avec $ 0 \leq \nu \leq 1$.

On peut \'etudier une variante des m\'eandres o\`u le nombre de routes est
laiss\'e libre~; les routes ne doivent pas se croiser, ni elles-m\^emes, ni
entre elles (mais peuvent \^etre embo\^{\i}t\'ees) et le nombre total de
ponts sur la rivi\`ere est toujours fix\'e a $n$.  Ce probl\`eme est alors
exactement \'equivalent~\cite{dfgg3} \`a celui d'une marche al\'eatoire de
$2n$ pas sur une demi-droite, le marcheur commen\c{c}ant et finissant \`a
l'extr\'emit\'e.  Chaque marche correspond alors \`a une configuration de
m\'eandres et r\'eciproquement.  Il y a alors $C_n$ marches possibles de ce
type, o\`u les $C_n$ sont les nombres de Catalan $(2n)! / (n+1)! / n!$.  On
en d\'eduit, pour ce mod\`ele soluble, que $R=4$, $\gamma = 3/2$ et $\nu =
1/2$ qui est l'exposant du mouvement Brownien.  Mais en imposant qu'il n'y
ait qu'une seule route, la nature du probl\`eme change compl\`etement et
les r\'esultats exacts sont rares.  En particulier, les valeurs ci-dessus
changent et n'ont pas encore \'et\'e calcul\'ees exactement.

On peut aussi d\'efinir d'autres quantit\'es comme la forme moyenne des
m\'eandres ou leur aires.  Plus de d\'etails sont donn\'es dans la
Ref.~\cite{og}.

%%%%%%%%%%%%%%%%%%%%%%%%%%%%%%%%
\section{D\'enombrement exact}
%%%%%%%%%%%%%%%%%%%%%%%%%%%%%%%%

  Pour construire les m\'eandres de taille $n+1$ \`a partir des m\'eandres
de taille $n$ de fa\c{c}on syst\'ematique, plusieurs mani\`eres sont
possibles.  Malheureusement, aucune d'entre elles ne donne de r\'ecurrence
directement entre le nombre total de m\'eandres $M_{n+1}$, et $M_n$~: il
faut \'etudier s\'epar\'ement chacun des $M_n$ m\'eandres.  Dans cette
section, nous d\'ecrivons une m\'ethode~\cite{dfgg1,dfgg3}, qui a \'et\'e
programm\'ee sur un ordinateur parall\`ele, le Cray-T3D.  Comme elle est a
la base de l'algorithme Monte-Carlo, nous la pr\'esentons en d\'etail.

  Cette m\'ethode consiste \`a ajouter un pont sur la partie la plus
\`a gauche de la rivi\`ere et \`a modifier la route pour la faire
passer par ce nouveau pont.  Pour que ce changement soit minimal, seule une
arche {\em ext\'erieure} est modifi\'ee (une arche est dite {\em
ext\'erieure} quand aucune autre ne l'entoure).

\begin{figure}
  \centering\leavevmode
  \epsfbox{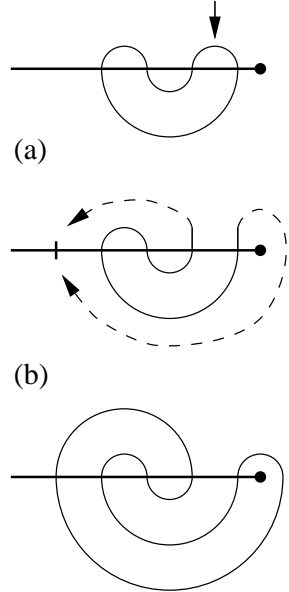}
  \caption{\em
    Un m\'eandre {\em (fils)} de taille $n+1$ est construit \`a partir
    d'un {\em p\`ere} de taille $n$ avec une arche ext\'erieure
    marqu\'ee de la mani\`ere suivante~: (a) Ajouter un pont sur la
    partie gauche de la rivi\`ere.  Couper l'arche ext\'erieure
    marqu\'ee et tirer sur ses deux extr\'emit\'es libres. (b) Refermer
    cette arche du cot\'e oppos\'e en passant par le nouveau pont (et
    \'eventuellement en contournant la source par la droite).  Le
    processus est inversible~: (b) Ouvrir la route \`a
    l'emplacement du pont le plus \`a gauche (a) Prendre les deux
    extr\'emit\'es libres et les refermer du cot\'e oppos\'e pour
    former une arche ext\'erieure.
  }
  \label{recur}
\end{figure}

  Prenons un m\'eandre de taille $n$ (le {\em p\`ere}) et choisissons
(ou marquons) une de ses arches ext\'erieures.  Par le processus
d\'ecrit Fig.~\ref{recur}, un m\'eandre de taille $n+1$ (le {\em fils})
est construit.  Le p\`ere a autant de fils que d'arches ext\'erieures.
En inversant le processus, chaque m\'eandre de taille $n+1$ a un et un
seul p\`ere.  C'est donc une correspondance un-\`a-un entre, d'une
part les m\'eandres de taille $n+1$, et d'autre part, ceux de taille
$n$ avec une arche ext\'erieure marqu\'ee.

\begin{figure}
  \centering\leavevmode
  \epsfbox{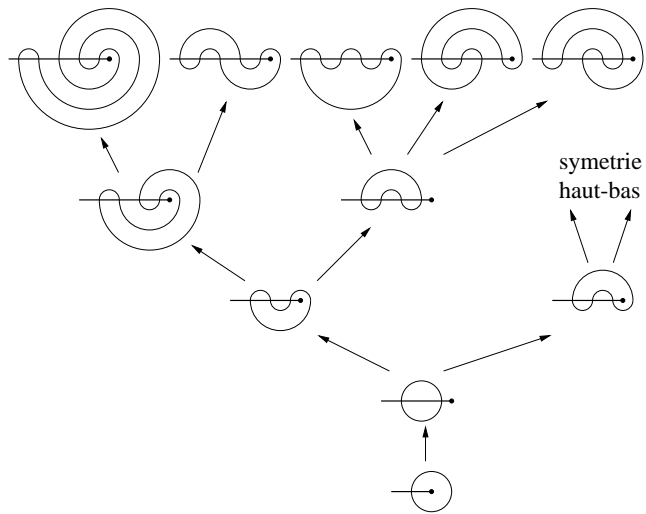}
  \caption{\em
    L'arbre des m\'eandres jusqu'\`a $n=5$.  Pour $n \geq 4$, seule une
    moiti\'e est dessin\'ee, l'autre \'etant obtenue par la
    sym\'etrie haut--bas.  Ainsi, $M_1 = 1$, $M_2 = 1$, $M_3 = 2$, $M_4
    = 4$ et $M_5=10$.  Chaque fl\`eche repr\'esente un processus
    ``p\`ere--fils''. 
  }
  \label{arbre}
\end{figure}

Le point de d\'epart de la r\'ecurrence est l'unique m\'eandre de taille
1.  En appliquant $n-1$ fois le processus, tous les m\'eandres de taille
$n$ peuvent \^etre construits.
Comme d\'ecrit Fig.~\ref{arbre}, l'ensemble des m\'eandres s'organise
comme un {\em arbre}.  La racine, au niveau 1, est le m\'eandre de
d\'epart $n=1$.  Chaque branche entre un n{\oe}ud au niveau $n$ et un
autre n{\oe}ud au niveau $n+1$ repr\'esente une relation entre un p\`ere
de taille $n$ et un de ses fils de taille $n+1$.  A l'exception de la
racine $n=1$, chaque m\'eandre (ou n{\oe}ud) a plusieurs arches
ext\'erieures, donc plusieurs fils (ou branches).  Leur nombre d\'epend
de la forme pr\'ecise du m\'eandre p\`ere et varie entre 2 et $n/2+1$.
Il s'agit donc d'un arbre {\em d\'eterministe} mais dont le nombre de
branches d\'epend du n{\oe}ud consid\'er\'e.

  Si l'on veut \'enum\'erer exactement les $M_n$ m\'eandres de taille $n$,
la seule m\'ethode que nous connaissons est de construire cet arbre
jusqu'au niveau $n$.  En particulier, nous n'avons pas trouv\'e de relation
de r\'ecurrence directement entre les $M_n$.  Le nombre de fils pour chaque
m\'eandre a une distribution qui semble erratique et le seul moyen de la
conna\^{\i}tre est l'examen de ses arches.  Aussi, pour calculer $M_n$, le
travail \`a fournir est proportionnel aux $M_n$, qui croissent
exponentiellement~: les limites de l'ordinateur sont vite atteintes.

\begin{figure}
  \centering\leavevmode
  \epsfxsize=\largeurFig
  \epsfbox{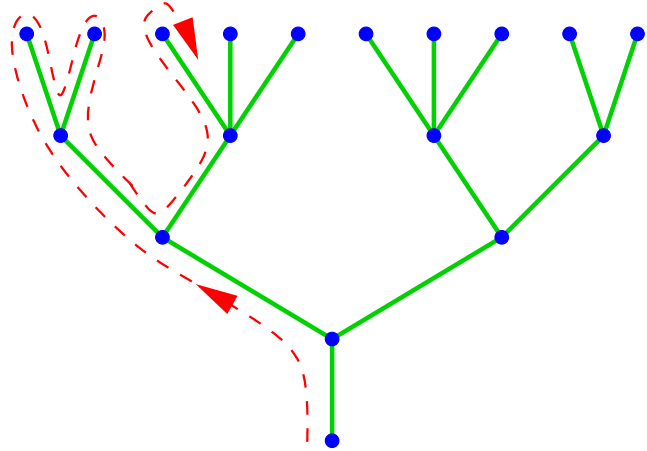}
  \caption{\em
    Algorithme de l'\'ecureuil m\'eticuleux~: il visite toutes les
    feuilles de la gauche vers la droite.
  }
  \label{meticul}
\end{figure}

Pour cela,
on coupe donc l'arbre au niveau $n$.  On appelle alors {\em feuilles}
les n{\oe}uds de niveau $n$ (qui repr\'esentent des m\'eandres de
taille $n$).  L'algorithme, sch\'ematis\'e Fig.~\ref{meticul},
consiste \`a visiter toutes les feuilles de la gauche vers la droite,
\`a la mani\`ere d'un ``\'ecureuil m\'eticuleux''.  Les r\`egles sont
les suivantes. (a) L'\'ecureuil d\'emarre \`a la racine. (b) Lorsque
l'\'ecureuil se trouve sur un n{\oe}ud interm\'ediaire (qui n'est pas
une feuille), il grimpe dans la branche la plus \`a gauche qu'il n'a
pas encore visit\'ee.  S'il a d\'ej\`a visit\'e toutes les branches,
il descend d'un niveau. (c) Lorsque l'\'ecureuil se trouve sur une
feuille, il descend d'un niveau.

Le lecteur peut se convaincre que ces r\`egles d\'ecrivent bien une
visite compl\`ete de l'arbre.  Bien sur, lorsque l'\'ecureuil se
trouve sur un n{\oe}ud, il mesure des quantit\'es int\'eressantes, par
exemple le nombre d'enroulements.  Ces mesures sont cumul\'ees et
trait\'ees \`a la fin de l'\'enum\'eration.

\begin{figure}
  \centering\leavevmode
  \epsfbox{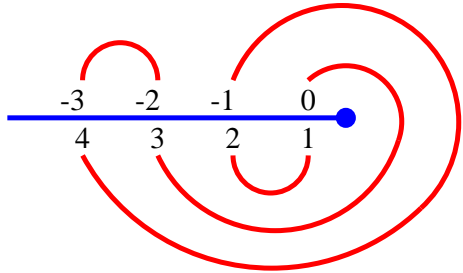}
  \caption{\em
    Pour ce m\'eandre de taille $4$, l'arche $(-1,4)$ est
    repr\'esent\'ee par $A(-1) = 4$ et $A(4) = -1$.  Au total,
    $A(-3:4)$ = $(-2$, $-3$, $4$, $3$, $2$, $1$, $0$, $-1$).
  }
  \label{arche}
\end{figure}

D'un point de vue Fortran, plusieurs repr\'esentations des m\'eandres
sont possibles.  Dans celle que nous avons finalement utilis\'ee,
chaque arche est cod\'ee par les num\'eros des deux ponts qu'elle
relie.  Il faut cependant distinguer les deux rives.  Aussi, les
ponts, vus de la rive sup\'erieure, sont num\'erot\'es n\'egativement
de $-n+1$ \`a 0, de la gauche vers la droite.  Vus de la rive
inf\'erieure, ils sont num\'erot\'es positivement de 1 \`a $n$, de la
droite vers la gauche.  Le pont le plus proche de la source porte les
num\'eros 0 et 1~; le plus \'eloign\'e les num\'eros $-n+1$ et $n$.
Ainsi, le syst\`eme des $n$ arches est cod\'e par le tableau d'entiers
$A(-n+1:n)$ o\`u $A(i)$ est le num\'ero du pont reli\'e par une arche
au pont $i$.  Voir Fig.~\ref{arche} pour un exemple.

Ce codage est loin d'\^etre optimal pour l'utilisation de la
m\'emoire.  On pourrait par exemple se contenter d'un tableau de $2n$
bits, o\`u le bit $i$ vaut 1 (resp. 0) si le pont $i$ est reli\'e \`a
un pont de num\'ero sup\'erieur (resp. inf\'erieur).  Cependant ce
codage permet de casser et de fusionner des arches efficacement.
En effet, lorsque l'\'ecureuil monte, une arche $(j,A(j))$ est
cass\'ee et donne deux arches limit\'ees par le nouveau pont~: $(-n,j)$
et $(A(j),n+1)$.  Lorsque l'\'ecureuil redescend, les deux arches
$(-n+1, A(-n+1))$ et $(A(n),n)$ fusionnent pour donner l'arche
$(A(-n+1),A(n))$, l'arche suivante \`a casser \'etant \'eventuellement
celle commen\c{c}ant \`a $j=A(n)+1$.

\begin{table*}
\begin{verbatim}
Integer, Parameter :: nmax = 20 ! hauteur maximale
Integer :: A(-nmax+1:nmax) = 0  ! representation en arches
Integer :: n = 1                ! hauteur de l'ecureuil
Integer :: j                    ! branche a visiter
Integer :: M(nmax)         = 0  ! compte les meandres de
                                !               taille n
A(0) = 1                        ! meandre de depart

Arbre : Do                      ! parcours de l'arbre
 M(n) = M(n) + 1                ! visite d'un nouveau noeud
 j = -n + 1                     ! arche la plus a gauche

 Do While (n==nmax .or. j==n+1) ! Descente :
   A(A(-n+1)) = A(n)            !  on fusionne les deux
   A(A(n))    = A(-n+1)         !      arches externes
   j = A(n) + 1                 !  arche suivante a casser
   n = n - 1
   If (n == 0) Exit Arbre       ! fin du parcours
 Enddo

 A(A(j)) = n+1                  ! Montee :
 A(n+1)  = A(j)                 !   l'arche (j) est cassee
 A(j)    = -n
 A(-n)   = j
 n = n + 1
Enddo Arbre

Print '(i3,i15)', (n, M(n), n = 1, nmax)
End
\end{verbatim}
    \caption{\em
       Programme Fortran qui compte les m\'eandres jusqu'\`a $n=nmax$.
    }
    \label{prog}
\end{table*}

Le programme Fortran, Table~\ref{prog}, compte donc les m\'eandres
jusqu'\`a $ n = nmax$ donn\'e.  Il est possible d'utiliser la sym\'etrie
haut-bas pour diviser le travail par deux.  On peut aussi se contenter de
ne construire les m\'eandres que jusqu'\`a la taille $n-1$ et le d\'ecompte
de leurs arches externes donne $M_n$.  On peut gagner $k$ \'etages
suppl\'ementaires, en utilisant des formules, exponentiellement
compliqu\'ees avec $k$, faisant intervenir pour chaque m\'eandre de taille
$n-k$, les nombres de ses arches externes, de ses arches de profondeur 2,
\ldots, jusqu'\`a la profondeur $k$.  Le choix optimal est $k=4$ ou 5.

Ce programme est clairement anti-vectoriel.  Pour donner un ordre de
grandeur, avec $n=20$, $M_{20} = 102511418$ est obtenu en 31 secondes
sur un Sun avec UltraSparc \`a 360 MHz.  Il met 70 secondes sur une
machine vectorielle~: le Fujitsu VPP-300 du Cea-Grenoble.  Pour
\'etudier les $n$ les plus grands possible avec cet algorithme (qui
est le seul dont on dispose), il ne reste plus que la voie du {\em
parall\'elisme massif}, o\`u l'ordinateur dispose de nombreux processeurs
calculant ind\'ependamment et interconnect\'es par un r\'eseau rapide.

  Pour cela, une taille interm\'ediaire $n_1$ est choisie.  Dans une
premi\`ere phase, tr\`es courte, l'arbre est construit, sur chaque
processeur, jusqu'au niveau $n_1$.  On obtient donc $M_{n_1}$ ``petits''
m\'eandres, par exemple $M_{11} = 4210$ avec $n_1=11$.

\begin{figure}
  \centering\leavevmode
  \epsfbox{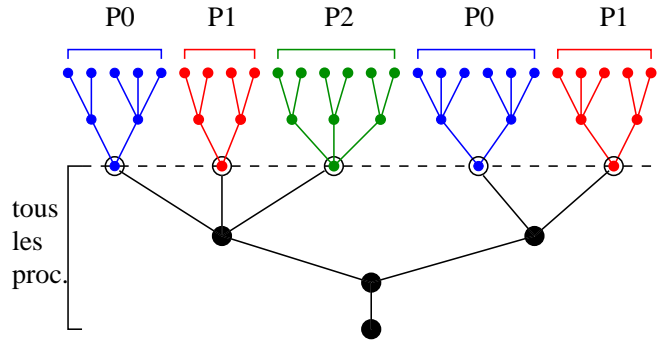}
  \caption{\em
     Parall\'elisation de l'algorithme de l'\'ecureuil m\'eticuleux~:
     dans une phase initiale tr\`es courte, tous les processeurs
     construisent l'arbre jusqu'\`a une hauteur $n_1$.  Les
     m\'eandres obtenus sont alors pris comme nouveaux points de
     d\'epart et le travail est r\'eparti de mani\`ere cyclique entre
     tous les processeurs, not\'es ici P0, P1 et P2.
  }
  \label{para}
\end{figure}

Dans une seconde phase, longue et massivement parall\`ele (voir
Fig.~\ref{para}), chacun de ces petits m\'eandres est consid\'er\'e
comme la {\em racine} d'un arbre, qui est alors un sous-arbre de
l'arbre global.  Chaque sous-arbre est trait\'e ind\'ependamment des
autres.  Sur une machine parall\`ele, le travail est r\'eparti entre
tous les processeurs, chacun traitant, \`a peu pr\'es, le m\^eme
nombre de sous-arbres.  A la fin, les mesures sont collect\'ees,
somm\'ees et trait\'ees par un seul processeur.

Pour un sous-arbre donn\'e, le travail \`a faire est proportionnel \`a
son nombre de n{\oe}uds~: il fluctue sensiblement d'un sous-arbre \`a
l'autre.  Mais comme chaque processeur a plusieurs dizaines de sous-arbres
\`a traiter, on constate que les fluctuations se moyennent et
l'\'equilibrage des charges entre processeurs est r\'ealis\'e \`a quelques
\% pr\`es.  Il n'est donc pas n\'ecessaire d'utiliser de m\'ethode plus
sophistiqu\'ee.

Pour programmer cela sur un Cray-T3D, les directives de compilation {\em
Craft}, fournies par le constructeur, ont \'et\'e utilis\'ees, car
l'algorithme s'y pr\^etait bien.  Cette extension de Fortran est du type
``data-parallel'' et a \'et\'e remplac\'ee depuis par Hpf.  Une directive
{\em barrier} synchronise tous les processeurs avant la collecte des
r\'esultats, qui se fait par le biais d'un tableau {\em partag\'e}, o\`u
chaque processeur y \'ecrit le r\'esultat de son d\'enombrement.  Puis,
gr\^ace \`a des directives {\em master} et {\em end master}, seul un
processeur fait et traite les sommes.  La difficult\'e principale n'a donc
pas \'et\'e le codage en soi de l'algorithme dans le langage parall\`ele,
mais bien la mise au point d'une parall\'elisation efficace.

  \begin{table}
    \centering
    \begin{tabular}{|rr|rr|}
      \hline
      $n$ & $M_n$ & $n$ & $M_n$ \\
      \hline
      1  &            1 &  16 &       1053874 \\
      2  &            1 &  17 &       3328188 \\
      3  &            2 &  18 &      10274466 \\
      4  &            4 &  19 &      32786630 \\
      5  &           10 &  20 &     102511418 \\
      6  &           24 &  21 &     329903058 \\
      7  &           66 &  22 &    1042277722 \\
      8  &          174 &  23 &    3377919260 \\
      9  &          504 &  24 &   10765024432 \\
     10  &         1406 &  25 &   35095839848 \\
     11  &         4210 &  26 &  112670468128 \\
     12  &        12198 &  27 &  369192702554 \\
     13  &        37378 &  28 & 1192724674590 \\
     14  &       111278 &  29 & 3925446804750 \\
     15  &       346846 &     &               \\
      \hline
    \end{tabular}
    \caption{\em
      Les nombres $M_n$ de m\'eandres de taille $n$ pour $n \leq 29$,
      obtenus par \'enum\'eration exacte sur ordinateur parall\`ele.
    }
    \label{n29}
  \end{table}

  Les r\'esultat jusqu'\`a $n=29$, donn\'es Table~\ref{n29}, ont
\'et\'e obtenus en 163 heures r\'eparties sur 128 processeurs du
Cray-T3D du Cea-Grenoble en 1995.  On voit que les $M_n$ croissent
exponentiellement.  Apr\`es extrapolation pour $n = \infty$, on en
d\'eduit que $ R = 3.50(1) $ et $ \gamma \approx 2$.

  Si la puissance des ordinateurs continue de cro\^{\i}tre
exponentiellement au fil des ann\'ees, comme le temps de calcul est
proportionnel aux $M_n$ qui croissent eux-aussi exponentiellement, le
mieux qu'on puisse esp\'erer avec cette m\'ethode de d\'enombrement
exact est un gain d'une nouvelle taille (de $n$ \`a $n+1$) tous les
deux ans environ.

%%%%%%%%%%%%%%%%%%%%%%%%%%%%%%%%%
\section{M\'ethode Monte-Carlo}
%%%%%%%%%%%%%%%%%%%%%%%%%%%%%%%%%

  Nous avons vu dans la section pr\'ec\'edente que la m\'ethode
d'\'enum\'eration exacte \'etait vite limit\'ee car le nombre $M_n$ de
m\'eandres \`a construire cro\^{\i}t exponentiellement avec la taille
$n$.  L'id\'ee g\'en\'erale des m\'ethodes Monte-Carlo est de
remplacer l'analyse compl\`ete de tous les cas possibles par la
s\'election {\em al\'eatoire} d'un sous-ensemble.  Celui sera
repr\'esentatif si la moyenne obtenue avec ces \'echantillons est
proche du vrai r\'esultat (qui est inconnu).  Cette moyenne est
al\'eatoire et fluctue d'une exp\'erience num\'erique \`a l'autre, ces
fluctuations \'etant alors empiriquement mesurables.

Pour le probl\`eme des m\'eandres, comme on ne sait construire
efficacement un m\'eandre de taille $n+1$ qu'\`a partir d'un m\'eandre
de taille $n$, nous allons nous servir de l'arbre d\'ecrit plus haut.

Nous allons d'abord introduire la m\'ethode qui utilise {\em un}
\'ecureuil Monte-Carlo.  Malheureusement cette m\'ethode n'est pas
efficace car ses fluctuations statistiques croissent tr\`es
vite avec $n$.  La parade consiste alors \`a utiliser une grande
population d'\'ecureuils, m\'ethode qui, de plus, est
parall\'elisable.

\subsection {Un \'ecureuil Monte-Carlo}

La vie de l'\'ecureuil Monte-Carlo commence \`a la racine de l'arbre
(\`a $n=1$).  Il grimpe de mani\`ere al\'eatoire dans l'arbre.  A
chaque niveau $n$, il se trouve sur un n{\oe}ud de l'arbre (qui
repr\'esente un m\'eandre de taille $n$) et fait un certain nombre de
mesures.  Il grimpe au niveau $n+1$ en choisissant au hasard de
mani\`ere uniforme l'une des $b_n$ branches partant de ce n{\oe}ud.
L'\'ecureuil s'arr\^ete \`a une certaine hauteur $n_{max}$
fix\'ee \`a l'avance.

\begin{figure}
  \centering\leavevmode
  \epsfbox{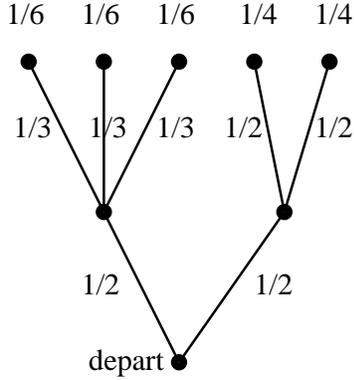}
  \caption{\em
     La probabilit\'e (nombres du haut) d'atteindre une feuille
     donn\'ee est le produit des probabilit\'es d'embranchement
     (nombres interm\'ediaires) rencontr\'ees \`a chaque noeud.
     Elle n'est pas uniforme.
  }
  \label{e1}
\end{figure}

Comme repr\'esent\'ee Fig.~\ref{e1}, la probabilit\'e que l'\'ecureuil
atteigne au niveau $n$ un n{\oe}ud donn\'e n'est pas uniforme.  Elle vaut
$1/q_n$ o\`u 
  \begin{equation}
     q_n = b_1.b_2\ldots b_{n-1}
  \end{equation}
est le produit des  nombres de branches $b_i$ que l'\'ecureuil a
rencontr\'es \`a chaque n{\oe}ud lors de son
ascension, et qu'il peut donc calculer.

Ce processus, c'est \`a dire une ascension Monte-Carlo d'un seul
\'ecureuil, est consid\'er\'e comme {\em une} simulation.  De nombreuses
simulations ind\'epen\-dantes sont r\'ealis\'ees et les mesures sont
moyenn\'ees.  Pour cela, il faut les pond\'erer avec le poids $q_n$.
Ainsi, chaque m\'eandre contribue au r\'esultat avec probabilit\'e $1/q_n$
et, le cas \'ech\'eant, le poids $q_n$, donc de mani\`ere uniforme.  En
particulier, l'esp\'erance math\'ematique $\langle q_n \rangle$, d\'efinie
comme la moyenne sur toutes les simulations possibles, vaut $M_n$.

  Malheureusement, cette m\'ethode ne marche pas en pratique pour $n$
grand.  En effet, bien que la loi de probabilit\'e de chaque $b_i$
soit r\'eguli\`ere, la loi des $q_n$, produit d'un grand nombre de
$b_i$, n'est pas {\em auto-moyennante}~: dans la limite $n$ grand,
avec probabilit\'e 1, les $q_n$ obtenus par simulation sont bien plus
petits que l'esp\'erance math\'ematique $\langle q_n \rangle$.  En
effet, le th\'eor\`eme de la limite centrale s'applique \`a $\ln q_n$,
qui est auto-moyennant car somme des $\ln b_i$, mais pas \`a $q_n$.
Plus pr\'ecis\'ement, des \'ev\'enements exponentiellement rares avec
$n$ contribuent de mani\`ere exponentiellement grande \`a $\langle q_n
\rangle$.  Dans une simulation, la moyenne des $q_n$ observ\'es sera
domin\'ee par ces \'ev\'enements rares et les fluctuations seront
larges.  Il faut donc un nombre exponentiellement grand (avec $n$) de
simulations.  Dans la pratique, on ne peut pas d\'epasser $n \approx
30$ ou $35$.

\subsection {M\'ethode Monte-Carlo multi-\'ecureuil}

La m\'ethode pr\'ec\'edente est g\'en\'eralis\'ee en utilisant une
population de $S$ \'ecureuils, o\`u $S$ est un param\`etre fix\'e \`a
l'avance.  Il est plus simple de le choisir parmi la liste des $M_n$~: $S =
M_{n_0}$.  Ainsi, au d\'epart, chaque n{\oe}ud au niveau $n_0$ est occup\'e
par un \'ecureuil.  Dans ce travail, $n_0=17$ et la sym\'etrie
haut-bas a \'et\'e utilis\'ee pour r\'eduire la population \`a $S =
M_{17}/2 = 1\,664\,094$ \'ecureuils.

\begin{figure}
  \centering\leavevmode
  \epsfbox{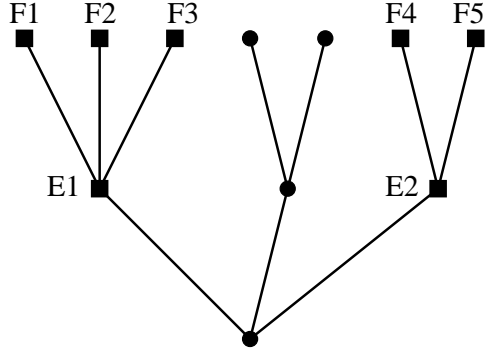}
  \caption{\em
    La m\'ethode Monte-Carlo multi-\'ecureuil est ici sch\'ematis\'ee
    avec $S=2$ \'ecureuils (E1 et E2).  Leurs $S'=5$ fils (F1 \`a F5)
    sont \'enum\'er\'es.  Pour garder $S$ constant, ils sont
    d\'ecim\'es et seulement $S=2$ fils, choisis au hasard, survivent.
    Ce processus est it\'er\'e jusqu'\`a atteindre la hauteur voulue
    l'arbre.
  }
  \label{e2}
\end{figure}

Les $S$ \'ecureuils \'evoluent, de mani\`ere synchrone, du niveau $n$ au
niveau $n+1$ par le processus suivant, sch\'ematis\'e Fig.~\ref{e2}.
Consid\'erons l'\'ecureuil d'index $i$ sur un n{\oe}ud au niveau $n$,
connect\'e \`a $b_i$ n{\oe}uds du niveau $n+1$.  Il se reproduit et donne
$b_i$ fils, chacun d'entre eux vivant sur un de ces $b_i$ n{\oe}uds
sup\'erieurs.  Tous les \'ecureuils se reproduisent en m\^eme temps et le
nombre total de fils $S' = \sum_{i=1}^S b_i$ est calcul\'e.  Le taux de
reproduction moyen $B_n = S'/S$ est une estimation du rapport
$M_{n+1}/M_n$.  Pour emp\^echer la croissance exponentielle de la
population (et donc des besoins en m\'emoire et en Cpu), celle-ci est
gard\'ee constante~: seulement $S$ parmi les $S'$ fils survivent.  Le choix
est fait au hasard de mani\`ere uniforme.  La probabilit\'e de survie d'un
fils est donc $1/B_n$.  Cette d\'ecimation est l'\'etape Monte-Carlo de
l'algorithme.  Puis, le processus s'applique \`a nouveau aux fils~; il est
it\'er\'e jusqu'\`a atteindre une hauteur $n_{max}$ fix\'ee \`a l'avance.
Cela constitue {\em une} simulation.  De nombreuses simulations
ind\'ependantes sont r\'ealis\'ees~: les moyennes et leurs barres d'erreur
associ\'ees sont alors calcul\'ees.

On voit que le cas particulier $S=1$ n'est rien d'autre que la
m\'ethode pr\'ec\'edente \`a {\em un} \'ecureuil.  A l'oppos\'e, la
limite $S=\infty$ est \'equivalente \`a la m\'ethode d'\'enum\'eration
exacte car on ne supprime pas d'\'ecureuil.

Mais quelque soit $S$, comme pour $S=1$, la probabilit\'e qu'un n{\oe}ud
soit visit\'e par un \'ecureuil n'est pas uniforme et il y a toujours un
biais.  Par exemple, les n{\oe}uds qui ont peu de fr\`eres, de
cousins, \ldots, ont plus de chance d'\^etre visit\'es.  On peut
montrer~\cite{og} que ce biais est exactement corrig\'e si les mesures sont
pond\'er\'ees, pour chaque simulation, par
\begin{equation}
  q_n = B_{n_0}\ldots B_{n-2}.B_{n-1},
\end{equation}
produit des facteurs de d\'ecimation.

A priori, cette m\'ethode souffre des m\^emes d\'efauts que la
pr\'ec\'edente, car le poids $q_n$ est le produit de nombreux $B_i$
al\'eatoires, donc non auto-moyennant quand $n$ devient grand.  Mais,
l'am\'elioration capitale est que la distribution de $B_i$ devient tr\`es
piqu\'ee autour de $M_{i+1}/M_i$ quand la population $S$ est grande.  En
effet, $B_i$ \'etant le nombre de fils en moyenne pour $S$ \'ecureuils, ses
fluctuations sont faibles et d'ordre $O(1/\sqrt{S})$.  Pour le dire
autrement, dans l'expression de $q_n$, la moyenne sur les $S$
\'ecureuils est faite au niveau des $B_i$, donc {\em avant} leur
produit.  Les fluctuations de $q_n$ croissent certes toujours
exponentiellement, mais avec un taux tr\`es faible si $S$ est grand.
Par exemple, pour $S = 1\,664\,094$ et $n = 400$, la largeur $\sigma$
de la distribution des $q_n$ est de $12 \%$ avec quelques \'ev\'enements
rares allant jusqu'\`a trois fois la moyenne.

Comment choisir $S$ et $n$ ?  Na\"{\i}vement, on a envie de simuler les
m\'eandres de taille $n$ la plus grande possible.  Mais, pour r\'ealiser
$N_s$ simulations ind\'epen\-dantes avec $S$ \'ecureuils jusqu'\`a la taille
$n$, le besoin de m\'emoire d'ordinateur est d'ordre $O(n.S$) et celui de
temps Cpu d'ordre $O(n^2.S.N_s)$.  Avec $S$ suffisamment grand, les
fluctuations statistiques sont presque gaussiennes et d'ordre
$O(1/\sqrt{S.N_s})$.  Comme $n$ est toujours limit\'e, nous extrapolerons
pour \'etudier la limite $n$ infini.  L'allocation de temps d'ordinateur
\'etant fix\'ee par ailleurs, nous nous sommes limit\'es \`a $n\leq 400$,
pour bien extrapoler avec une meilleure statistique.  Finalement, \`a
produit $S.N_s$ constant, il vaut mieux choisir $S$ le plus grand possible
permis par la m\'emoire de l'ordinateur, pour diminuer le probl\`eme des
fluctuations rares et grandes de $q_n$.
 
L'algorithme a \'et\'e ici d\'ecrit en terme d``\'ecureuils''.  Mais
ceux-ci repr\'e\-sentent des m\'eandres par l'interm\'ediaire d'un
tableau d'entiers $A(-n+1:n)$.  En particulier, l'op\'eration qui
transforme un \'ecureuil en son fils, consiste \`a casser une arche
externe, donc \`a manipuler ce tableau.  Cette op\'eration n'est pas
vectorisable et s'av\`ere \^etre la partie qui consomme le plus de Cpu
dans l'algorithme.  Comme dans toute simulation Monte-Carlo, la
qualit\'e statistique des r\'esultats est limit\'ee par le temps Cpu.
Il est alors capital de parall\'eliser l'algorithme.

La parall\'elisation la plus simple consiste \`a r\'ealiser des simulations
totalement ind\'ependantes sur chacun des processeurs.  Il suffit
d'initialiser diff\'e\-remment le g\'en\'erateur de nombres al\'eatoires d'un
processeur \`a l'autre, et d'organiser la collecte des mesures effectu\'ees
lors des simulations pour les traiter ult\'erieurement.  Ce type de
parall\'elisation n'utilise que tr\`es peu le r\'eseau car rien n'est
\'echang\'e entre les processeurs.  Cela peut m\^eme se concevoir sur un
r\'eseau h\'et\'erog\`ene de stations de travail, de mani\`ere asynchrone.

Mais on a vu ci-dessus qu'il \'etait tr\`es important de simuler la plus
grande population possible, plut\^ot que de multiplier les simulations avec
une population plus petite, la limite \'etant donn\'ee par la m\'emoire
disponible.  Pour cela, tous les processeurs doivent travailler de concert
sur la m\^eme population.  L'algorithme utilis\'e est d\'ecrit ci-apr\`es.

Dans une courte phase d'initialisation, les $S$ \'ecureuils sont
construits au niveau $n_0$ et sont r\'epartis de mani\`ere
\'equilibr\'ee parmi les processeurs. Chaque processeur ne s'occupe
que de son groupe d'\'ecureuils.

Lors d'une it\'eration principale, o\`u la population passe du niveau
$n$ au niveau $n+1$, chaque processeur \'enum\`ere les fils de ses
\'ecureuils, puis r\'ealise et somme ses mesures.  Les sommes globales sont
ensuite faites sur l'ensemble des processeurs et \'ecrites dans un fichier
par un seul processeur pour \^etre analys\'ees ult\'erieurement.  La
facteur de d\'ecimation $B_n$ doit \^etre d\'etermin\'e de mani\`ere
globale, et non pas localement sur chaque processeur.  Un processeur,
d\'eclar\'e ma\^{\i}tre, analyse le nombre de fils $F_p$ obtenus par chaque
processeur $p$, calcule $B_n$ et fixe en retour le nombre $F_p/B_n$ de fils
que chaque processeur doit garder.  La population globale est donc
maintenue exactement \'egale \`a $S$.  Cette phase demande une
synchronisation entre processeurs, peu de calculs et l'\'echange de petits
messages.  Puis vient la d\'ecimation proprement dite, r\'ealis\'ee de
mani\`ere ind\'ependante sur chacun des processeurs.  C'est l'\'etape qui
consomme le plus de Cpu car les fils survivants sont effectivement
construits \`a ce moment-l\`a.

Il est alors n\'ecessaire d'\'equilibrer les populations entre processeurs.
En effet, le nombre de fils fluctue. Lors de la synchronisation, tous les
processeurs doivent attendre celui qui a le plus d'\'ecureuils, car il
effectue plus de travail.  Sans y rem\'edier, ce probl\`eme empirerait au
fil des it\'erations.  Aussi, \`a la fin de chaque it\'eration, le
processeur ma\^{\i}tre recense les processeurs dont la population d\'epasse
de 5~\% la moyenne et leur fait transf\'erer la partie exc\'edentaire aux
processeurs les moins charg\'es.  Ainsi, les temps d'attente sur la
barri\`ere de synchronisation sont en moyenne de 5 \%.  En contrepartie, le
r\'eseau transf\`ere 5 \% de la m\'emoire de quelques processeurs.

Ce processus constitue {\em une} simulation.  Plusieurs simulations sont
effectu\'ees de mani\`eres ind\'ependantes, en un ou plusieurs ``runs''.
Les fichiers de mesures sont ensuite trait\'es par un autre programme.

Cette parall\'elisation a \'et\'e programm\'ee sur le Cray-T3E du
Cea-Grenoble, \'equip\'e de processeurs Dec-alpha \`a 375 MHz.  Il faut 13
Goctets de m\'emoire pour une population de $S = 1\,664\,094$ \'ecureuils,
soit 128 processeurs de 128 Moctets.  Pour r\'ealiser 8192 simulations, 8
jours ont \'et\'e n\'ecessaires, soit 24000 heures de Cpu.  Pour
parall\'eliser, la biblioth\`eque Shmem, fournie par le constructeur, a
\'et\'e utilis\'ee.  Elle est du type ``passage de message'', comme PVM ou
MPI.  Son int\'er\^et principal est sa simplicit\'e et son efficacit\'e.
Par contre, elle n'est pas portable.

Le rendement de la parall\'elisation est proche des 95~\%
th\'eoriques.  Aucune d\'egradation significative n'a \'et\'e vue,
m\^eme avec 128 processeurs.  En effet, les \'echanges entre
processeurs sont rares et de volume moyen~: ils ne posent aucun
probl\`eme pour le r\'eseau du T3E, qui est un point fort de cette
machine.  La performance du programme est plut\^ot
sensible aux probl\`emes d'acc\`es au cache et \`a la m\'emoire
locale~; un gain important a \'et\'e obtenu, lors des copies de tableaux
dans une m\^eme m\'emoire locale, gr\^ace aux registres ``E'', en
court-circuitant le Cpu et son cache.

On voit que le travail scalaire par processeur est proportionnel \`a son
nombre d'\'ecureuils~; par contre la taille des messages est
proportionnelle aux fluctuations, donc \`a la racine carr\'ee.  En
th\'eorie, la ``scalabilit\'e'' est parfaite, \`a condition d'augmenter la
population globale --- ce qui est justement int\'eressant --- quand on
augmente le nombre de processeurs, pour que la population par processeur ne
s'appauvrisse pas.

%%%%%%%%%%%%%%%%%%%%%%%%%%%%%%%%
\section{Quelques r\'esultats}
%%%%%%%%%%%%%%%%%%%%%%%%%%%%%%%%

On s'int\'eresse tout d'abord au comportement asymptotique du nombre de
m\'eandres $ M_n \sim c R^n / n^{\gamma} $ pour $n$ grand.  L'entropie
$\ln R$ peut \^etre estim\'ee avec
\begin{equation}
  L_n = \frac{1}{2} \ln \left( \frac{M_n}{M_{n-2}} \right),
\end{equation}
o\`u $n$ saute de 2 en 2 pour diminuer des effets sensibles de parit\'e
entre les $n$ pairs et impairs.  Comme on s'attend \`a $L_n \sim \ln R -
\gamma / n$ pour $n$ grand, en tra\c{c}ant $y = L_n$ en fonction de
$x=1/n$, on pourra estimer $\ln R$ (limite quand $x$ tend vers 0) et
$\gamma$ (pente asymptotique).

\begin{figure}
  \centering\leavevmode
  \epsfxsize=\largeurFig
  \epsfbox{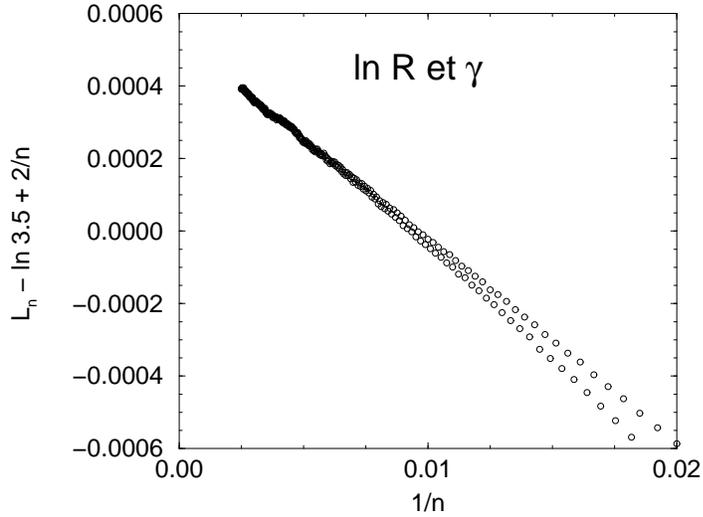}
  \caption{\em    
    Estimation Monte-Carlo de $L_n-\ln 3.5 + 2/n$
    en fonction de $1/n$ pour $n$ allant de $50$ \`a $400$.  La limite
    quand $x$ tend vers $0$ est $\ln (R/3.5)$, et la pente (n\'egative) est
    $2-\gamma$.  Les barres d'erreur ne sont pas dessin\'ees car elles sont
    toujours inf\'erieures \`a $10^{-5}$, donc plus petites que les symboles.
    Un effet de parit\'e entre les $n$ pairs et impairs est visible.  Une
    extrapolation lin\'eaire donne $R = 3.5019(2)$ et $\gamma = 2.056(10)$.
  }
  \label{lnr}
\end{figure}

Sur la Fig.~\ref{lnr}, l'estimation Monte-Carlo de $L_n-\ln 3.5 + 2/n$ est
trac\'ee en fonction de $1/n$ pour $n$ allant de 50 \`a 400.  La fonction
{\em lin\'eaire} $y = \ln 3.5 -2x$ a \'et\'e arbitrairement soustraite pour
r\'eduire l'amplitude de variation de $y$~; les quantit\'es int\'eressantes
$2-\gamma$ (pente r\'esiduelle), $\ln(R/3.5)$ (limite quand $x$ tend vers
0) et la courbure (d\'eviation au comportement lin\'eaire esp\'er\'e) sont
ainsi rendues plus visibles.  Malgr\'e cette amplification, la courbure
reste faible et rend raisonnable une extrapolation lin\'eaire, qui donne
\begin{equation}
  R = 3.5019(2) \;\;\; \mathrm{et} \;\;\; \gamma = 2.056(10),
\end{equation}
en contradiction avec les conjectures $7/2$ et 2.

\begin{figure}
  \centering\leavevmode
  \epsfxsize=\largeurFig
  \epsfbox{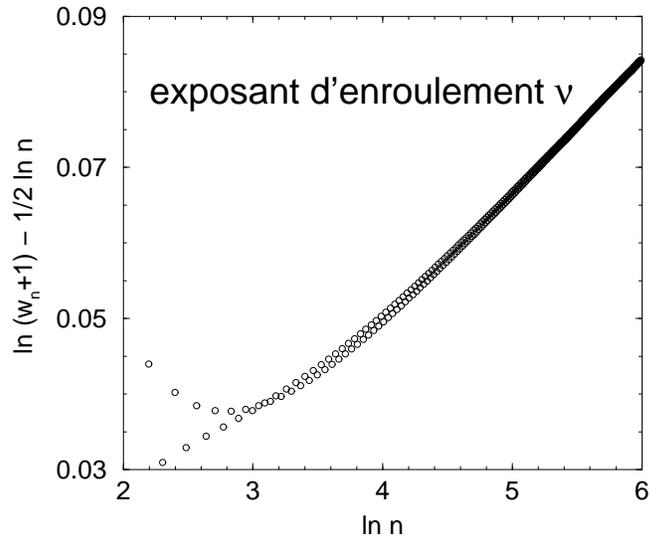}
  \caption{\em 
     Estimation Monte-Carlo de $\ln(w_n+1) - 1/2 \ln n$ en fonction de $\ln
     n$, pour $n$ entre $8$ et $400$. La pente est de $0.018$ et mesure $\nu -
     1/2$.  Les barres d'erreur ne sont pas dessin\'ees car inf\'erieures \`a
     $3.10^{-4}$.
  }
  \label{fnu}
\end{figure}

L'{\em exposant d'enroulement} $\nu$, d\'efini par le comportement
asymptotique du nombre moyen d'enroulements $w_n \sim n^{\nu}$, est
pr\'esent\'e Fig.~\ref{fnu}.  En tra\c{c}ant $\ln(w_n)$ en fonction de $\ln
n$, la pente asymptotique est une mesure de $\nu$.  Nous pr\'ef\'erons
utiliser $\ln(w_n+1)$ au lieu de $\ln(w_n)$, car on remarque que $w_n+1$
est moins sensible que $w_n$ aux effets de taille finie.  Comme la question
principale est de savoir si $\nu=1/2$ ou non, la fonction {\em lin\'eaire}
$y = x/2$ a \'et\'e soustraite.  Ainsi, la variation de $y$ est r\'eduite~;
$\nu-1/2$ et la courbure sont plus visibles.  On voit que cette derni\`ere
reste petite et une extrapolation lin\'eaire donne
\begin{equation}
  \nu = 0.518(2),
\end{equation}
incompatible avec la valeur brownienne 1/2.

D'autres r\'esultats sont pr\'esent\'es dans la R\'ef~\cite{og}, concernant
la loi de probabilit\'e du nombre d'enroulements et la forme moyenne des
m\'eandres, en particulier pr\`es de la source et de l'extr\'emit\'e
oppos\'ee.

%%%%%%%%%%%%%%%%%%%%%%
\section{Conclusion}
%%%%%%%%%%%%%%%%%%%%%%

Dans ce travail, nous avons simul\'e le repliement compact d'un objet
unidimensionnel, une bande de $n$ timbres-poste, mod\'elis\'ee par les
m\'eandres.  Jusqu'\`a pr\'esent, on ne conna\^{\i}t pas de formule donnant
le nombre de pliages possibles $M_n$ en fonction de $n$, et l'ordinateur
est donc une m\'ethode d'investigation privil\'egi\'ee.  Les algorithmes,
que nous avons d\'evelopp\'es et utilis\'es, reposent sur une
repr\'esentation de l'ensemble de tous les pliages possibles en un arbre de
forme irr\'eguli\`ere.

Avec un algorithme de visite syst\'ematique de tous les n{\oe}uds de
l'arbre jusqu'\`a une hauteur fix\'ee $n$, on peut mesurer exactement
toutes les propri\'et\'es.  Mais la nature m\^eme de l'arbre emp\^eche la
vectorisation, car il n'y a pas de longue boucle r\'eguli\`ere.  Par
contre, la possibilit\'e de le d\'ecouper en sous-arbres rend la
parall\'elisation efficace sur une machine \`a grand nombre de processeurs.
Mais le travail cro\^{\i}t exponentiellement avec $n$ et m\^eme un
ordinateur puissant ne permet pas de d\'epasser $n \sim 30$.

Avec un algorithme Monte-Carlo, on peut \'etudier des m\'eandres plus
grands.  Mais les fluctuations statistiques, inh\'erentes \`a ce type de
m\'ethode, sont ici particuli\`erement d\'efavorables.  Avec la mise au
point de la m\'ethode multi-\'ecureuil, les moyennes sont r\'ealis\'ees en
deux \'etapes~: la premi\`ere, entre \'ecureuils, a des fluctuations
gaussiennes, le cas le plus favorable. On cherche donc \`a simuler
simultan\'ement le plus grand nombre possible d'\'ecureuils, ce qui est
principalement limit\'e par la m\'emoire de l'ordinateur.  En
r\'epartissant la population sur de nombreux processeurs, la
parall\'elisation permet alors de b\'en\'eficier de beaucoup plus de
m\'emoire.  On obtient aussi une forte puissance en Cpu scalaire,
n\'ecessaire pour que les r\'esultats significatifs ne soient pas noy\'es
dans les fluctuations statistiques, ce qui a permis alors des simulations
jusqu'\`a la taille $n=400$.

On voit donc que les \'etudes d\'ecrites dans cet article doivent beaucoup
aux ordinateurs parall\`eles, les ordinateurs vectoriels \'etant
inefficaces pour ces algorithmes.  La parall\'elisation \'etant hors de
port\'ee d'un compilateur automatique, elle doit \^etre explicitement
programm\'ee, le ``passage de message'' s'av\'erant le mieux adapt\'e.

%%%%%%%%%%%%%%%%%%%%%%%%%%%%%
\subsection*{Remerciements}
%%%%%%%%%%%%%%%%%%%%%%%%%%%%%

Nous remercions L. Colombet, P. Di Francesco, E. Guitter et R. Napoleone
pour toutes les discussions enrichissantes et l'aide \`a la
parall\'elisation des programmes.

%%%%%%%%%%%%%%%%%%%%%%%%%%%%%

\end{document}